\documentclass[preprint,12pt]{elsarticle}

\usepackage{graphicx}
\usepackage{amssymb}

{}
{}
{}

\usepackage{multirow}
\usepackage[boxruled,vlined,linesnumbered]{algorithm2e}
\usepackage{xcolor}
\usepackage{graphicx}
\usepackage{placeins}
\usepackage[export]{adjustbox}

\journal{Journal Name}

\begin{document}

\begin{frontmatter}

\title{Code-Aware Combinatorial Interaction Testing}

\author{Bestoun S. Ahmed}
\address{Department of Mathematics and Computer Science, Karlstad University\\ 651 88 Karlstad, Sweden\\ email: bestoun@kau.se}

\author{Angelo Gargantini}
\address{Department of Management, Information and Production Engineering\\ University of Bergamo, Italy\\ email: angelo.gargantini@unibg.it}
\author{Kamal Z. Zamli}
\address{Faculty of Computer Systems and Software Engineering, University Malaysia Pahang\\ email:kamalz@ump.edu.my}

\author{Cemal Yilmaz}
\address{Faculty of Engineering and Natural Sciences, Sabanci University, Istanbul, Turkey\\ email: cyilmaz@sabanciuniv.edu }

\author{Miroslav Bures and Marek Szeles}
\address{Department of Computer Science, Faculty of Electrical Engineering \\Czech Technical University in Prague, Karlovo nam. 13, 121 35 Praha 2, Czech Republic\\ email: buresm3@fel.cvut.cz\\ email: marek.szeles@eforce.cvut.cz}

\begin{abstract}
Combinatorial interaction testing (CIT) is a useful testing technique to address the interaction of input parameters in software systems. In many applications, the technique has been used as a systematic sampling technique to sample the enormous possibilities of test cases. In the last decade, most of the research activities focused on the generation of CIT test suites as it is a computationally complex problem. Although promising, less effort has been paid for the application of CIT. In general, to apply the CIT, practitioners must identify the input parameters for the Software-under-test (SUT), feed these parameters to the CIT tool to generate the test suite, and then run those tests on the application with some pass and fail criteria for verification. Using this approach, CIT is used as a black-box testing technique without knowing the effect of the internal code. Although useful, practically, not all the parameters having the same impact on the SUT. This paper introduces a different approach to use the CIT as a gray-box testing technique by considering the internal code structure of the SUT to know the impact of each input parameter and thus use this impact in the test generation stage. We applied our approach to five reliable case studies. The results showed that this approach would help to detect new faults as compared to the equal impact parameter approach.

\end{abstract}

% \begin{keyword}
% Internet of Things \sep IoT \sep Quality Assurance \sep Model-Based Testing

% \end{keyword}

\end{frontmatter}

\section{Introduction}

Modern software systems are increasingly getting large in terms of size, functionality, input parameters, and configurations. The different interaction (combinations) of input parameters or configurations may cause faults while running the system. Kuhn \textit{et al.} \cite{Kuhn2009} reported many interaction faults in mission-critical and other software systems. Combinatorial interaction testing (CIT) (sometimes called $t-way$ testing, where $t$ is the interaction strength) offers a sampling strategy that can effectively and efficiently sift out only fewer interactions to equate the otherwise impossible exhaustive testing of all interactions. To generate the CIT test suite, covering arrays were used to sample the inputs of the SUT based on the input interaction coverage criteria. This approach has been used for input combination and configuration testing as a black-box approach. To sample the input parameters in the combinatorial input testing, each row in the covering array represents a complete set of input to the software-under-test (SUT), while each row in configuration testing represents a configuration setting of the SUT \cite{Yilmaz:2013}. In both cases, the testing process is done as black-box testing where the internal code is not used during the test generation.

%%%%%%%%%%%%%%% Bestoun 7/2/2019 --- 2

While the black-box approach with the CIT is useful in many applications, not all the test cases are equally effective in finding faults, especially in the input parameter CIT testing \cite{TZOREFBRILL2019,Simos:2019}. The current test case generation algorithms do not consider the parameter impact on the SUT. In fact, CIT is generally used as a black-box testing strategy. Practically, not all the input parameters have the same impact on the internal code structure of the SUT. One way of defining the effect of a parameter is to consider those parameters that cover more lines of code to have more impact. Considering those parameters to generate or refactor the generated test cases may lead to a more practical CIT.

In this paper, we argue that CIT could be more useful when it also considers the internal code of the SUT. Here, we propose to mix black-box testing with white box testing (i.e., gray-box testing approach) where we extend CIT by reusing the information that is coming from the code. We introduce our code-aware approach to generate combinatorial interaction test suites. The approach relies on the generation of more effective test suites by considering the internal code structure of the SUT through the feedback from the code coverage analysis. Here, the test generation process relies on a preassessment and analysis of the internal code structure of the SUT to know the sensitivity for each input parameter and thus knowing the impact of each one of them. By understanding this impact, then, instead of using uniform interaction strength among the input parameters, we put more focus on those impacted parameters by considering higher or even full interaction strength. Hence, we use mixed strength instead of uniform strength. Our aim is to generate more effective test cases (in term of better fault detection capability and new fault finding) by identifying those input parameters which practically affect the SUT.

The rest of this paper is organized as follows. Section \ref{Motivation} gives the literature background of the CIT and essential concepts of the testing and generation algorithms. Section \ref{method} illustrates our method in this paper including the analysis and testing procedures. Section \ref{EmpiricalInvestigation} illustrates our empirical investigation of the code-aware combinatorial interaction testing approach, including the results and discussion. Section \ref{threatsToValidity} discusses possible threats to validity. Finally, Section \ref{Conclusion} gives the concluding remarks of the paper.

%%%%%%%%%%%%%%% Bestoun 7/2/2019 --- 3

\section{Basic Concepts and Literature}\label{Motivation}

\subsection{Basic Concepts}
Theoretically, the combinatorial test suite depends on a well-known mathematical object called Covering Array (CA). To represent a test suite, each row in the CA presents a test case and each column represents an input parameter of the SUT. Formally, a $CA_{\lambda}(N ; t, k, v)$ is an $N \times k$ array over $(0, . . . , v - 1)$ such that every $B=\lbrace b_0, ..., b_{t-1} \rbrace $ is $\lambda$-covered and every $N \times t$ sub-array contains all ordered subsets from $v$ values of size $t$ at least $\lambda$ times, where the set of column $B=\lbrace b_0, ..., b_{t-1} \rbrace \supseteq \lbrace0, ..., k-1 \rbrace$ \cite{Ahmed2015,Hartman2005}. In this case, each tuple is to appear at least once in a CA.

%%%%%%%%%%%%%%% Bestoun 8/2/2019 --- 4

When the number of component values varies, this can be handled by Mixed Covering Array (MCA). A $MCA (N; t, k, (v_1, v_2,..., v_k))$, is an $N \times k$ array on $v$ values, where the rows of each $N \times t$ sub-array covered and all $t$ interactions of the values from the $t$ columns occur at least once. For more flexibility in the notation, the array can be presented by $MCA (N; t, v_1^{k_1} v_2^{k_2} .. v_k^k)$.

In real-world complex systems, the interaction strength may vary between the input parameters. In fact, the interaction of some input parameters may be stronger than other parameters. Variable strength covering array (VSCA) is introduced to cater for this issue. A $VSCA (N; t; k, v, (CA_1, ..., CA_k))$ represents $N \times p$ MCA of strength $t$ containing vectors of $CA_1$ to $CA_k$, and a subset of the $k$ columns each of strength $ > t$ \cite{Ahmed:2011:VSI, BestounBioInspired}. Also, practically, not all the inputs are interacting and having an impact on each other. Some parameters may not interact at all. Here, it is not necessary to cover all the interactions of the parameter. Presenting those parameters even in one test case would be enough.

%%%%%%%%%%%%%%% Bestoun 8/2/2019 --- 5

\subsection{Literature and Motivation}

CIT used the aforementioned mathematical objects as a base for the testing strategy of different applications. A wide range of applications appeared in the literature. Mainly, CIT used in software testing and program verification. There are many applications in this direction, for example, fault detection, and characterization \cite{Colbourn2018,AvocadoBestoun2019}, graphical user interface testing (GUI) \cite{Yuan:2011}, model-based testing and mutation testing \cite{Bures2017, Tao2019}. There are many more applications of CIT in software testing.  Comprehensive surveys about these applications can be found in \cite{Kuhn:2013:ICT,BestouIEEAcess,Nie:2011}. The concepts of CIT also finds its way to other fields rather than software testing. For example, it has been used in the satellite communication testing, hardware testing \cite{Borodai1992}, advance material testing \cite{Schubert2004}, dynamic voltage scaling (DVS) optimization \cite{Sulaiman2013}, tuning the parameter of fractional order PID controller \cite{Sahib2016}, and gene expression regulation \cite{Shasha1590,Deaconnature2019}.

%%%%%%%%%%%%%%% Bestoun 8/2/2019 --- 6

In most of the applications, the combinatorial interaction test suite is generated by establishing a coverage criterion. Here, the coverage criterion is to cover the $t-tuples$ of the input parameter at least once to generate the CA. A few researchers in the literature considered some other input attributes during the generation in addition to the $t-tuple$ coverage criteria. For example, Yilmaz \cite{Yilmaz:2013} considered the test case-specific interaction constraints which are test cases related to the configurations. Demiroz and Yilmaz \cite{Demiroz:2016} also introduced the cost-aware covering arrays that generate the test cases based on a given cost function by modeling the actual cost of testing in addition to the standard $t-tuple$ coverage criteria.  

%%%%%%%%%%%%%%% Bestoun 8/2/2019 --- 7

Almost, in all applications, the SUT is considered as a black-box system by the generation tool. However, in practice, not all the parameters have the same impact on the internal code of the SUT. Logically, each value of the input parameters may have a different impact on the program. Taking into account the code coverage as an example, those parameters' tuples with higher code coverage may have a higher chance to cause faults. In this paper, we have considered this situation to take the CIT in a gray-box testing approach by analyzing the program internals.

\begin{figure}
\centering
\includegraphics[width= 4 in]	{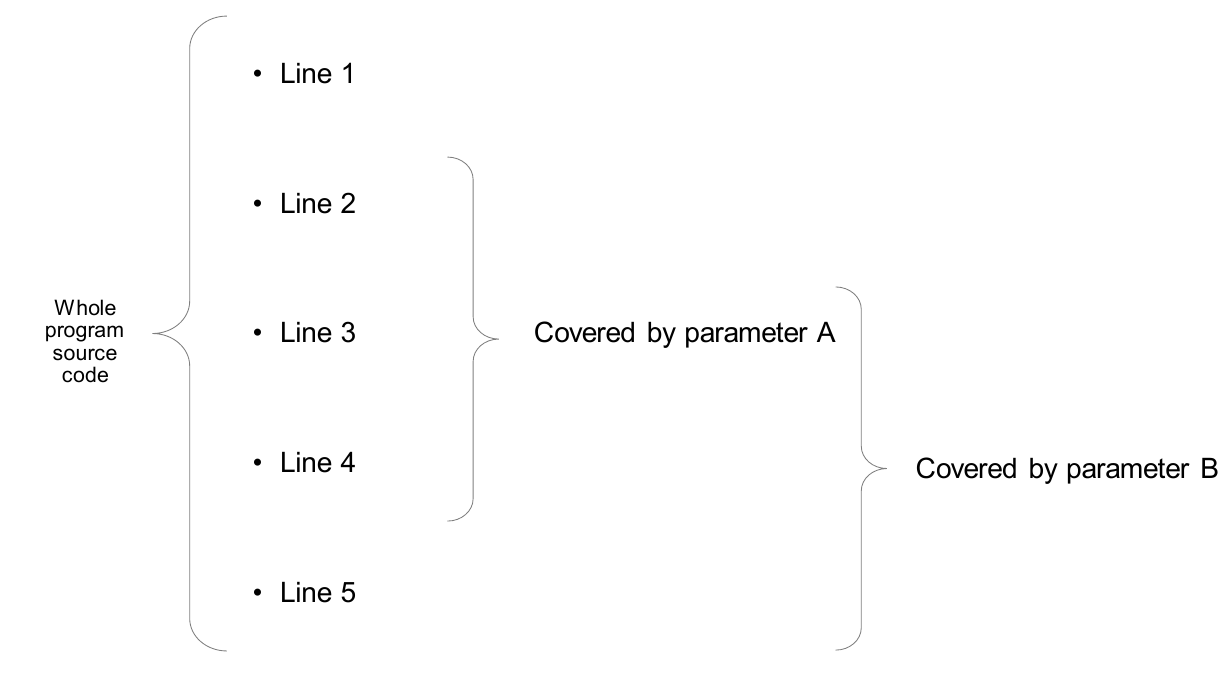}
\caption{Overlapping parameter code coverage demonstration on a hypothetical 5-line program}
\label{OverlapExample}

\end{figure}

In practice, the sum of individual parameter code coverage may exceed 100\%. This is because two (or more) parameters may impact the same code snippet both at the same time. To illustrate this overlapping issue, let's assume a hypothetical 5-line program and parameters A and B as shown in Figure \ref{OverlapExample}. Both parameters cover three LOC -- parameter A covers lines 2 to 4 and parameter B covers lines 3 to 5. Thus, both parameters cover 60\% of the code. Combined, however, they obviously cannot cover 120\% of the code, but they only cover 4 out of 5 lines together, i.e., 80\%. One line, or 20\% of the code, in this case, is not covered by any parameter. In practice, these lines would be exceptions, loggers, unused code and sometimes also comments.

\begin{figure}
\centering
\includegraphics[width= 4 in, frame]	{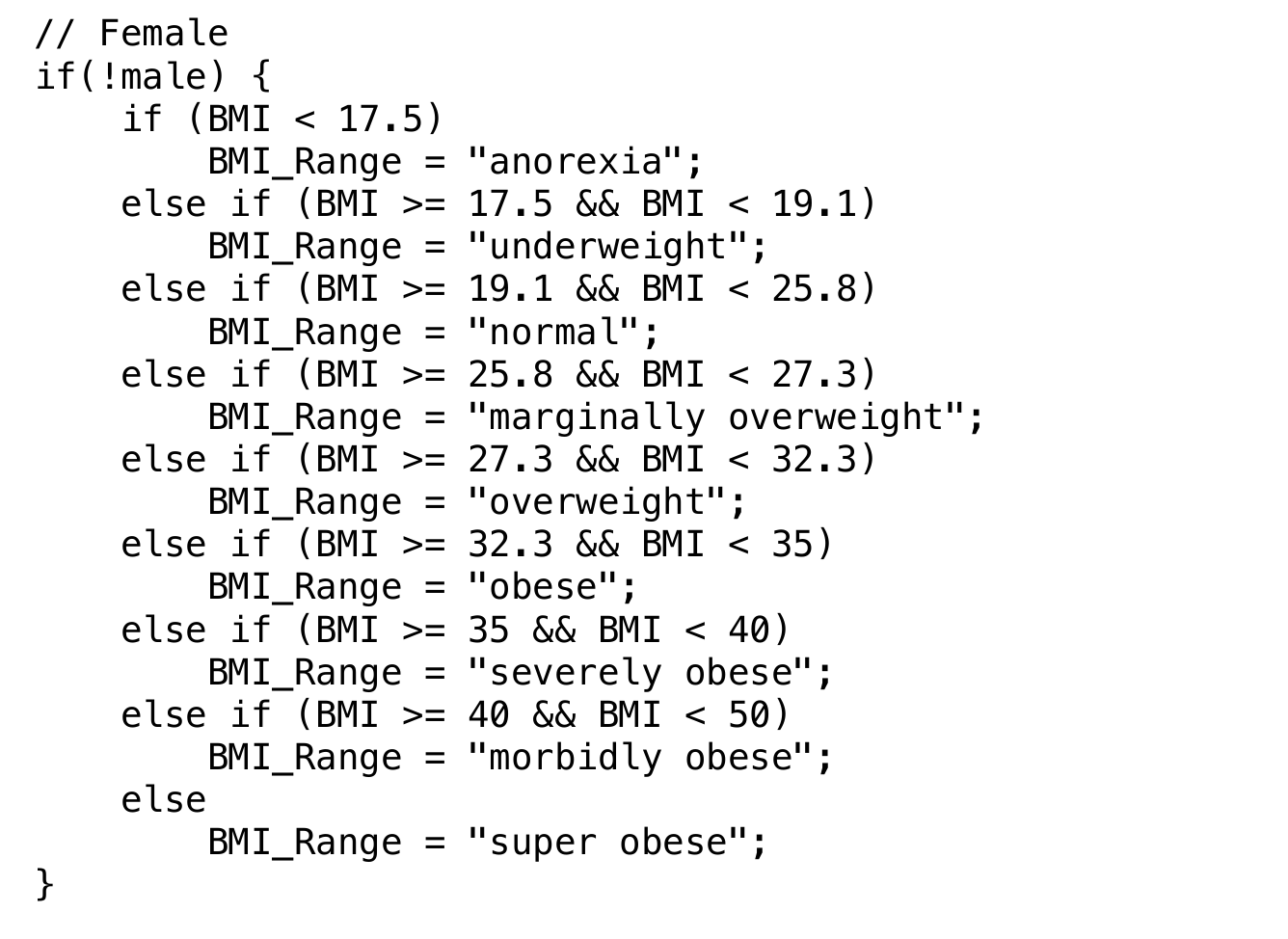}
\caption{As sample code snippet for the Overlapping parameter code coverage}
\label{SampleCodeOverlapExample}

\end{figure}

As a practical example, note in Figure \ref{SampleCodeOverlapExample} the code snippet from the case study (BMI calculator) that we used in this paper. The whole code snippet is affected by the value of the Boolean variable "male"  -- it is executed only if it is false (i.e., female). However, within it, there is an if-else switch based on the value of the double variable BMI -- based on which, different lines of code are activated, effectively changing the value of BMI\_Range. Thus, most lines in this code snippet are affected by both the variables BMI and male.

%%%%%%%%%%%%%%% Bestoun 8/2/2019 --- 8

\section{Method}\label{method}
For this study, we only consider those software systems which can be tested using the CIT approach. Here, the test suite can be represented as a CA with more than two input parameters. Each parameter has different values. We hypothesize that if we measure the parameter strength based on code coverage, we can figure out how much that input parameter has an impact on the SUT. Larger impact means that the parameter covers more lines of SUT code and thus it is more prone to failures. Hypothetically, those parameters may have better fault detection rates. Let $X$ and $Y$ be two parameters of a program $P$ and let their code coverage $C_X > C_Y$.

\begin{figure}
\centering
\includegraphics[width= 4 in]	{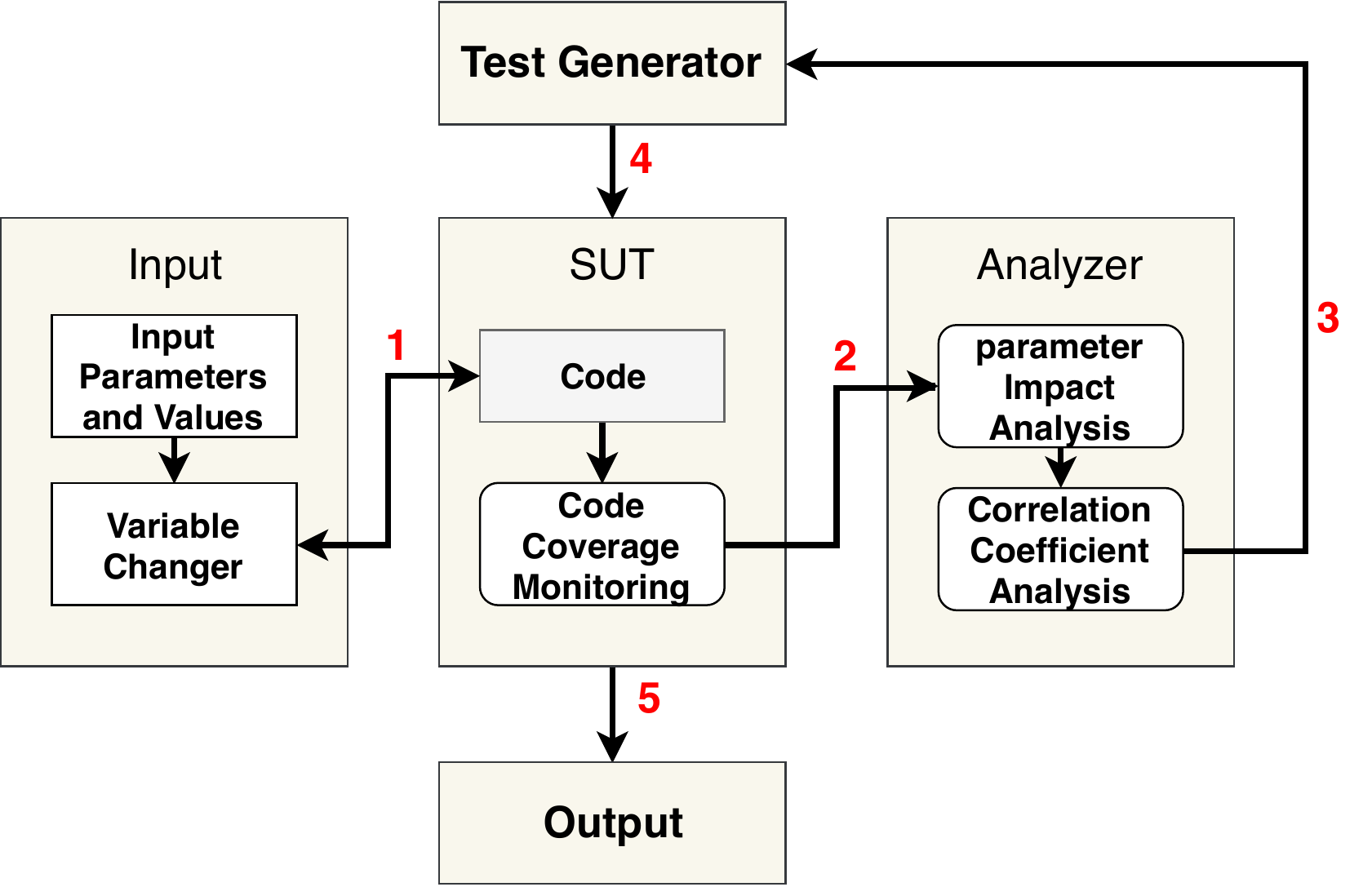}
\caption{Experimental method}
\label{Experimetnal_Method}

\end{figure}

To analyze the parameter impact of each SUT, we have followed several systematic experimental steps. Figure \ref{Experimetnal_Method} shows these steps.

First (step 1 in Fig. \ref{Experimetnal_Method}) we have identified the parameters and the values and we have called the code with a set of tests. The detail of the test generation method is illustrated in subsection \ref{TestGenerationMethod}. Here, we used the conventional $t-wise$ method to generate the set of tests. To avoid the omission of parameters' values effect on each other, we have tried two different possibilities and combinations of the values. First, we attempted to measure the coverage while we varied the values of a specific parameter and made the rest constant. Second, we also tried different combinations of values. Both approaches lead to the same conclusion as we are using the deviation at the end. Measuring the code coverage when we vary the value of a specific parameter (by the Variable Changer in Figure \ref{Experimetnal_Method}) while we make the rest constant will lead to conclude the impact of that parameter. However, in some situation, the code coverage may be affected by some other values of the other parameters. We considered this situation also to measure the code coverage of two parameters (i.e., pairwise) while making the others constant. Hence, for fair experimental results, we considered all the possibilities of the parameter effects on each other. 

%%%%%%%%%%%%%%% Bestoun 11/2/2019 --- 9

\begin{figure}
\centering
\includegraphics[width= 2 in]	{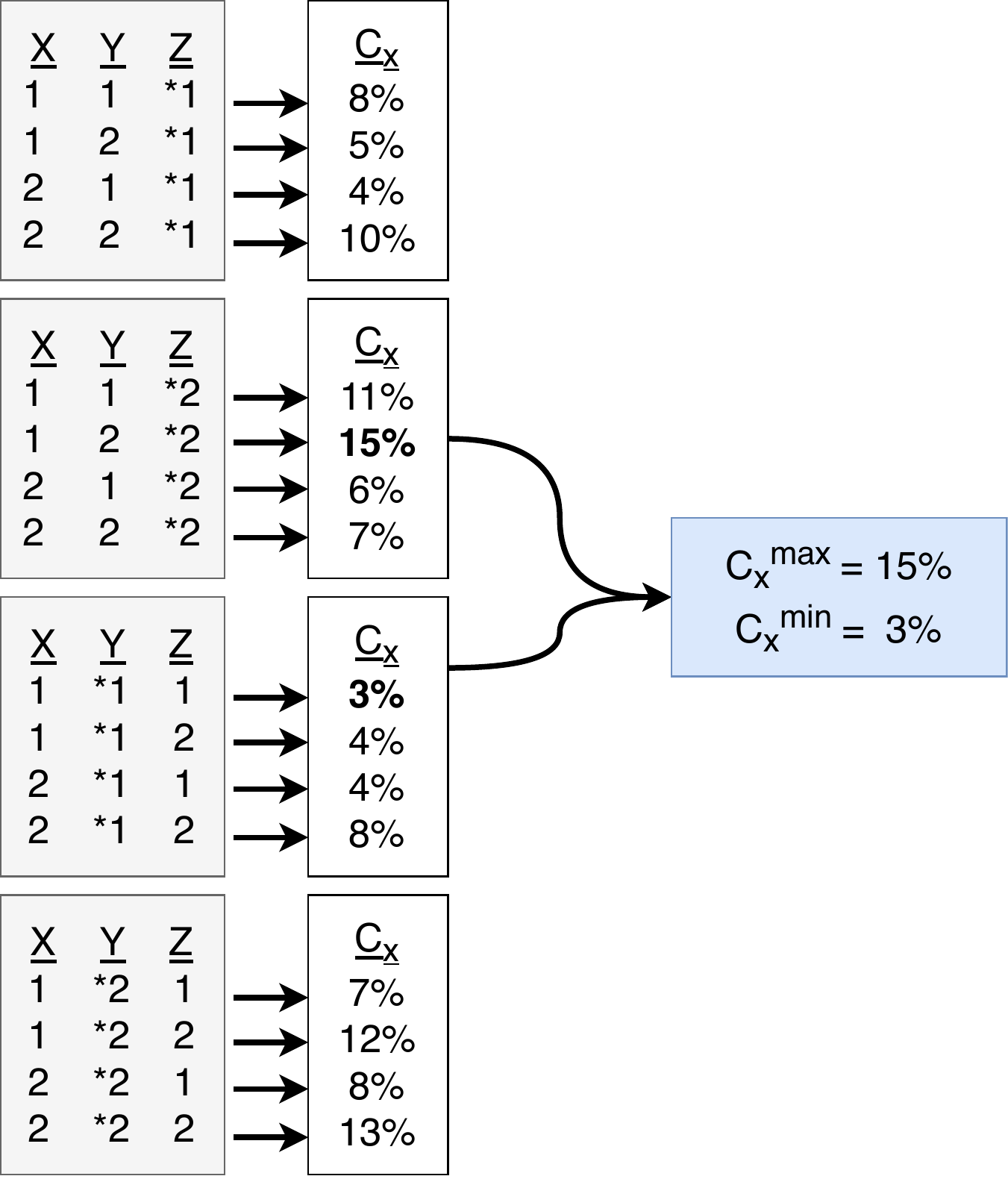}
\caption{Code coverage measurement method}
\label{Impact_Analysis}
\end{figure}

By measuring the code coverage (step 2 in Fig. \ref{Experimetnal_Method}), we can calculate each parameter impact by measuring the deviation of code coverage when entering different parameter values. For a fair experimental procedure, we used the best and worst code coverage situation as maximum and minimum code coverage. To measure the code coverage, we used an automated scripting framework to monitor and measure the impact of each input parameter. To calculate the impact of a parameter, we calculate the code coverage deviation as in Eq \ref{Eq1}.

\begin{equation}
I_p=C_p^{max}-C_p^{min}\label{Eq1}
\end{equation}

%%%%%%%%%%%%%%% Bestoun 11/2/2019 --- 10

where $I_p$ is the parameter impact, $C_p^{max}$ is the maximum code coverage of parameter $p$, and $C_p^{min}$ is the minimum code coverage of parameter $p$. The Eq \ref{Eq1} can be used to calculate the impact of pairwise parameters also, which can be used for the correlation coefficient later in Eq \ref{Eq:3}. To better illustrate the code coverage analysis process, Figure \ref{Impact_Analysis} shows a simplified example for a system with three input parameters (X, Y, and Z) where each parameter has two values (1 and 2). For illustration, we assign random numbers to the code coverage for each test case. The aim here is to measure the impact of parameter X. Here, the maximum and minimum code coverage of X are 15\% and 3\% respectively. Parts of these experimental records can be used mutually for the impact analysis of Y and Z also. Here, we consider the pairwise possibilities for the parameter values to consider all the parameters between X and Y, while assigning a constant value to Z (values with "*" sign). To avoid the omission of other values, we also change the constant values for the parameters for each pairwise experiment.

Using this approach to know the impact of the parameters, we can calculate the interaction weight by calculating the correlation of the parameters. For parameters X and Y, the correlation would be as in Eq.\ref{Eq:3}:

\begin{equation}
Corr(X,Y) = \frac{{{I_{XY}}}}{{{I_X} * {I_Y}}}\label{Eq:3}
\end{equation}

where $I_X$ and $I_Y$  are the impact of the parameters X and Y respectively, and the $I_{XY}$ is the impact of both of them. As mentioned previously, we have undertaken a careful code coverage monitoring to isolate the effects.  

Now, with this correlation, for each tuple of parameters, we create an $n \times n$ square matrix, where n is the number of parameters, as in Table \ref{CorrelationTable}.

\begin{table}

\caption{Conceptual model of parameter correlation mapping}\label{CorrelationTable}
\begin{centering}
\scriptsize
\begin{tabular}{|c|c|c|c|c|}
\hline 
parameter & $P_{1}$ & $P_{2}$ & ... & $P_{n}$\tabularnewline
\hline 
\hline 
$P_{1}$ & - & $Corr(P_{1},P_{2})$ & ... & $Corr(P_{1},P_{n})$\tabularnewline
\hline 
$P_{2}$ & $Corr(P_{2},P_{1})$ & - & ... & $Corr(P_{2},P_{n})$\tabularnewline
\hline 
... & ... & ... & - & ...\tabularnewline
\hline 
$P_{n}$ & $Corr(P_{n},P_{1})$ & $Corr(P_{n},P_{2})$ & ... & -\tabularnewline
\hline 
\end{tabular}
\par\end{centering}
\end{table}

%%%%%%%%%%%%%%% Bestoun 11/2/2019 --- 11

Using the data in Table \ref{CorrelationTable}, we can select and device the higher impacted tuples by assigning a partner of each parameter $p_i$, as in Eq.\ref{Eq:4}.

\begin{equation}
partner({p_i}) = \arg {\max _{{p_j}}}\;Corr({p_i},{p_j})\label{Eq:4}
\end{equation}

Note that since the matrix in Table \ref{CorrelationTable} is reflexive and diagonally mirrors itself, the same tuple will be present twice. For example, if the best partner for $p_1$ is $p_2$, then the best partner for $p_2$ is $p_1$. Therefore, this tuple will appear in the matrix both as $Corr({p_2},{p_1})$ and $Corr({p_1},{p_2})$. Both instances represent the same tuple however.

In practice, one can assign higher interaction strength to not only pairs of parameters, but also a set of three and more. Assigning higher interaction strength to a set of all parameters would result in no comparative change in strength impact between the individual parameter subsets. If one would decide to assign a higher strength to three or more parameters, it can be done in two ways. The first approach is to create a multi-dimensional matrix (n-dimensional for sets of n), and the methodology is the same. Alternatively, it is possible to spot a good set of three candidates in a two-dimensional matrix.

Due to the goal of this paper, we did not pay attention to the execution cost of the test suites. As mentioned previously, we aim to assess the effectiveness of our code-aware CIT approach via an experimental study. Based on this experimental study, these steps can be automated and abstracted to minimize the execution cost for the ease of use in the industry. The pseudo-code in Algorithm \ref{alg:automated-algorithm} describes those essential steps and how they can be executed. The pseudo-code may be used in the future to build a fully automated tool to fulfill these step successfully. We explained the details of the steps we followed in the experiments in the following sub-sections.

\begin{algorithm}

\KwIn {The source code of the SUT}

\KwOut {Program analysis report}

Identify the input parameters

\ForEach{Input parameter}{

\ForEach{Value of the parameter}{

Change the value

Run the Code and measure the code coverage
}

}

Analyze the parameter impact based on the code coverage cost

Computer the correlation coefficient for each parameter

Generate a set of t-wise test cases 

Produce an analysis report

Add higher interaction strength $t$ to subset of the input parameters 

Regenerate the test cases 

Execute the test suite and monitor the output

Produce the final analysis report

\caption{\label{alg:automated-algorithm} Code-aware CIT conceptual steps}

\end{algorithm}

\subsection{Mutant generation and fault seeding}

To analyze the effectiveness of our approach, we generate different types of mutants to be injected into the subjected programs for experiments. We used $\mu java$\footnote{https://cs.gmu.edu/~offutt/mujava/}, the classical java mutation tool, to generate the mutants. $\mu java$ is a mutation system for Java programs. It automatically generates mutants for both traditional mutation testing and class-level mutation testing. $\mu java$ can test individual classes and packages of multiple classes. Tests are supplied by the users as sequences of method calls to the classes under test encapsulated in methods in JUnit classes.

For the fault seeding, we have seeded 35 types of faults, as defined by the $\mu java$ documentation. Two levels of faults were used here, method-level faults and class-level faults. Mutants in the method-level fault seeding are based on changing operators within methods, or the complete statement alteration. In this analysis, 16 different types are considered. Table \ref{methodLevelFaults} shows those method-level faults and the correspondence abbreviations. Mutants based on class-level fault seeding are based on changing operators within methods, or the complete statement alteration. In this analysis, 29 different types are considered. Table \ref{ClassLevelFaults} shows those class-level faults and the correspondence abbreviations.

\begin{table}

\centering
\caption{Types and abbreviation of method-level faults used for the experiments}
\scriptsize
\begin{tabular}{ l l }
\hline 
Abbreviation & Meaning\tabularnewline
\hline 
\hline 
AOR & Arithmetic Operator Replacement\tabularnewline
\hline 
AOI & Arithmetic Operator Insertion\tabularnewline
\hline 
AOD  & Arithmetic Operator Deletion\tabularnewline
\hline 
ROR  & Relational Operator Replacement\tabularnewline
\hline 
COR  & Conditional Operator Replacement\tabularnewline
\hline 
COI  & Conditional Operator Insertion\tabularnewline
\hline 
COD  & Conditional Operator Deletion\tabularnewline
\hline 
SOR  & Shift Operator Replacement\tabularnewline
\hline 
LOR & Logical Operator Replacement\tabularnewline
\hline 
LOI &  Logical Operator Insertion \tabularnewline
\hline 
LOD  & Logical Operator Deletion\tabularnewline
\hline 
ASR  & Assignment Operator Replacement\tabularnewline
\hline 
SDL  & Statement DeLetion\tabularnewline
\hline 
VDL  & Variable DeLetion\tabularnewline
\hline 
CDL  & Constant DeLetion\tabularnewline
\hline 
ODL  & Operator DeLetion\tabularnewline
\hline 
\end{tabular}
\label{methodLevelFaults}

\end{table}

\begin{table*}[ht]
\caption{Types and abbreviation of class-level faults used for the experiments}
\label{ClassLevelFaults}
\begin{center}
\scriptsize
\begin{tabular}{lll}
\hline 
Feature & Abbreviation & Meaning\tabularnewline 
\hline 
\hline 
Encapsulation & AMC & Access modifier change\\
\hline 
    \multirow{8}{*}{Inheritance} & IHD & Hiding variable deletion\\
    & IHI  & Hiding variable insertion \\ %\cline{2-2} 
    & IOD  & Overriding method deletion \\ %\cline{2-2} 
    & IHI  & Hiding variable insertion \\
    & IOP  & Overriding method calling position change \\
%\cline{2-3} 
 & IOR  & Overriding method rename\tabularnewline
%\cline{2-3} 
 & ISI  & Super keyword insertion\tabularnewline
%\cline{2-3} 
 & ISD  & Super keyword deletion\tabularnewline
%\cline{2-3} 
 & IPC  & Explicit call to a parent\textquoteright s constructor deletion \\
    \hline
\multirow{10}{*}{Polymorphism} & PNC  & New method call with child class type\tabularnewline
%\cline{2-3} 
 & PMD  & Member variable declaration with parent class type\tabularnewline
%\cline{2-3} 
 & PPD  & Parameter variable declaration with child class type\tabularnewline
%\cline{2-3} 
 & PCI  & Type cast operator insertion\tabularnewline
%\cline{2-3} 
 & PCC  & Cast type change\tabularnewline
%\cline{2-3} 
 & PCD  & Type cast operator deletion\tabularnewline%\cline{2-3} 
 & PRV  & Reference assignment with other comparable variable\tabularnewline %\cline{2-3} 
 & OMR  & Overloading method contents replace\tabularnewline%\cline{2-3} 
 & OMD  & Overloading method deletion\tabularnewline%\cline{2-3} 
 & OAC  & Arguments of overloading method call change\tabularnewline
\hline     
\multirow{10}{*}{Java--specific features} & JTI  & This keyword insertion\tabularnewline
 & JTD  & This keyword deletion\tabularnewline
 & JSI  & Static modifier insertion\tabularnewline
 & JSD  & Static modifier deletion\tabularnewline
 & JID  & Java Member variable initialization deletion\tabularnewline
 & JDC & Java-supported default constructor deletion\tabularnewline
 & EOA  & Reference assignment and content assignment replacement\tabularnewline
 & EOC  & Reference comparison and content comparison replacement\tabularnewline
 & EAM  & Access or method change\tabularnewline
 & EMM  & Modifier method change\tabularnewline
\hline 
\end{tabular}
\end{center}
\end{table*}

%%%%%%%%%%%%%%% Bestoun 11/2/2019 --- 12

\subsection{Test case generation}\label{TestGenerationMethod}

To generate the test cases, we used ACTS3.0\footnote{https://bit.ly/2s2IajU}, the automated CIT tool that contains many algorithms to generate the combinatorial interaction test suites. ACTS is a well-known combinatorial interaction test generation tool that supports the generation of different interaction strength ($1\leq t \leq 6$). The tool provides both command line and GUI interfaces. The tool also offers the flexibility to address the interaction strength for different sets and subsets of input parameters, (i.e., mixed strength and variable strength interaction). To avoid the randomness of the test generation algorithms and to assure fair experiments statistically, we used the deterministic test generation algorithms in ACTS. Since ACTS is a combinatorial testing tool, it requires values of the different input parameters to be tested. Such values were produced by using arbitrary classes of presumed equivalence determined by the data type of a parameter and its default value. For example, the values typically tested were the original default value of a parameter, 0, negative input and factor multiples of the value which significantly affected the output, or runtime behavior. To assure for reasonable testing runtimes (as described in Section \ref{EmpiricalInvestigation}, hundreds of mutants were tested for each scenario), no more than five classes of equivalence were used for each parameter.

%%%%%%%%%%%%%%% Bestoun 11/2/2019 --- 13

\subsection{Parameter impact analysis}

Here, we adopted code coverage monitoring tools in our environment to analyze the code coverage during the first round of test execution. By analyzing the code coverage, we know the impact of the parameters with respect to the code. We adopted Sofya\footnote{http://sofya.unl.edu/} Java bytecode analysis tool for the code coverage analysis. Sofya is a tool designed to provide analysis capabilities for Java programs by utilizing the Bytecode Engineering Library (BCEL) to manipulate the class files. We have also used the base JetBrains\footnote{https://www.jetbrains.com/} IntelliJ IDE for the code coverage measurement. The first round takes several iterations depending on the input parameters and the value of each one of them. As previously mentioned, the program examine each parameter to identify $C_p^{max}$ and $C_p^{min}$ in Eq.\ref{Eq1}.

\section{Empirical investigation}\label{EmpiricalInvestigation}
In this section, we illustrate our empirical investigation. Here, five software subjects were used as case studies for the investigation. We first describe these case studies and then the fault injection and code analysis procedures. During this empirical investigation, we aim to answer three main research questions (RQ)s: 

\begin{itemize}
\item RQ1: To which extent the input parameters affect the internal code of the SUT?
\item RQ2: How the input parameters are correlated?
\item RQ3: How does the new approach improve the effectiveness of fault detection in the CIT? 
\end{itemize}

%%%%%%%%%%%%%%% Bestoun 11/2/2019 --- 14

\subsection{Case studies}

We chose five Java-based subject programs for the experiments. Two programs from Software-artifact Infrastructure Repository (SIR)\footnote{http://sir.unl.edu/portal/index.php} and three programs from other reliable software repositories. These programs are well-known experimental programs used for experimental purposes in other research studies. We choose "Replicated workers", "Groovy", "Body calculator\footnote{https://bit.ly/2rrUolM}", Searching\footnote{https://bit.ly/2HVxBKJ}, and Mortgage\footnote{https://bit.ly/2HRZSBP} for the case studies.

The replicated workers is an implementation of a standardized Replicated Workers problem. In some parallel algorithms, the number of specific computing tasks is not known in advance. To control the allocation of the tasks of the replicated workers, a work pool is used. The replicated workers program has 342 lines of code. The groovy snippet is a part of the core for Apache Groovy, a multi-faceted language for the Java platform. The groovy program has 361 lines of code. The body calculator application/applet is a medical program used for fat percentage, body mass index, Basal Metabolic Rate, Ideal Weight, and Calorie Intake. The body calculator program has 910 lines of code. The Searching program implements several sorting algorithms and sorts of a randomly generated input matrix based on a set of constraining input parameters. The searching program has 1084 lines of code. The Mortgage program is a GUI implementation of a mortgage amortization table calculator based on several user-defined inputs. The Mortgage program has 1045 lines of code.

%%%%%%%%%%%%%%% Bestoun 11/2/2019 --- 15

\subsection{Experimental procedure}
To follow the methodology given in Section \ref{method}, we have created two sets of test cases for each case study -- one reference set without parameter impact, and another set using our approach by considering the parameter impact. To know the effectiveness, we have seeded faults into the programs. We run each program with and without seeded faults to kill the mutants by recognizing the differences. Based on our approach, we then regenerated the test suites by putting interaction strength on those parameters which are correlated more to each other by considering the correlation coefficient, as illustrated in Table \ref{CorrelationTable}. For example, if the correlation coefficient of two parameters A and B is equal to 0.9, it means they are highly correlated to each other, we put the interaction strength on both of them. On another hand, if the correlation coefficient between the two parameters C and D is equal to 0.1, it means that the correlation is too low between them, we assign a "don't care" value for both of them. Hence, they will be presented in the test suites, but we don't care about the full coverage of their values. We run the newly generated test suite to identify the differences in fault detection.

%%%%%%%%%%%%%%% Bestoun 11/2/2019 --- 16

\subsection{The seeded faults}

\begin{figure}
\centering
\includegraphics[width= 4 in]	{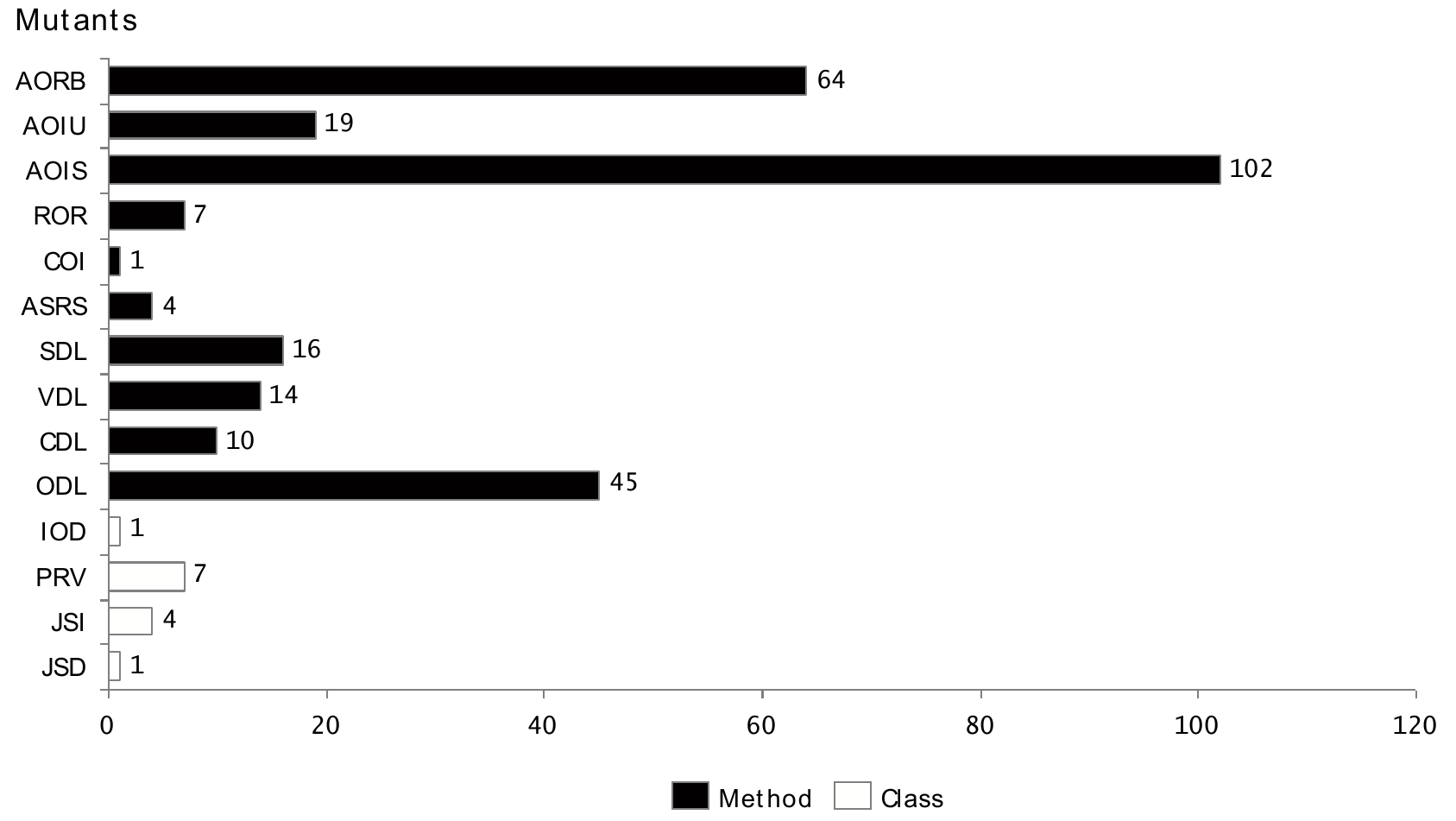}
\caption{Type and number of seeded faults into the Replicated Worker program}
\label{ReplicatedWorkerFaultsSeeded}

\end{figure}

\begin{figure}
\centering
\includegraphics[width= 3.3 in]	{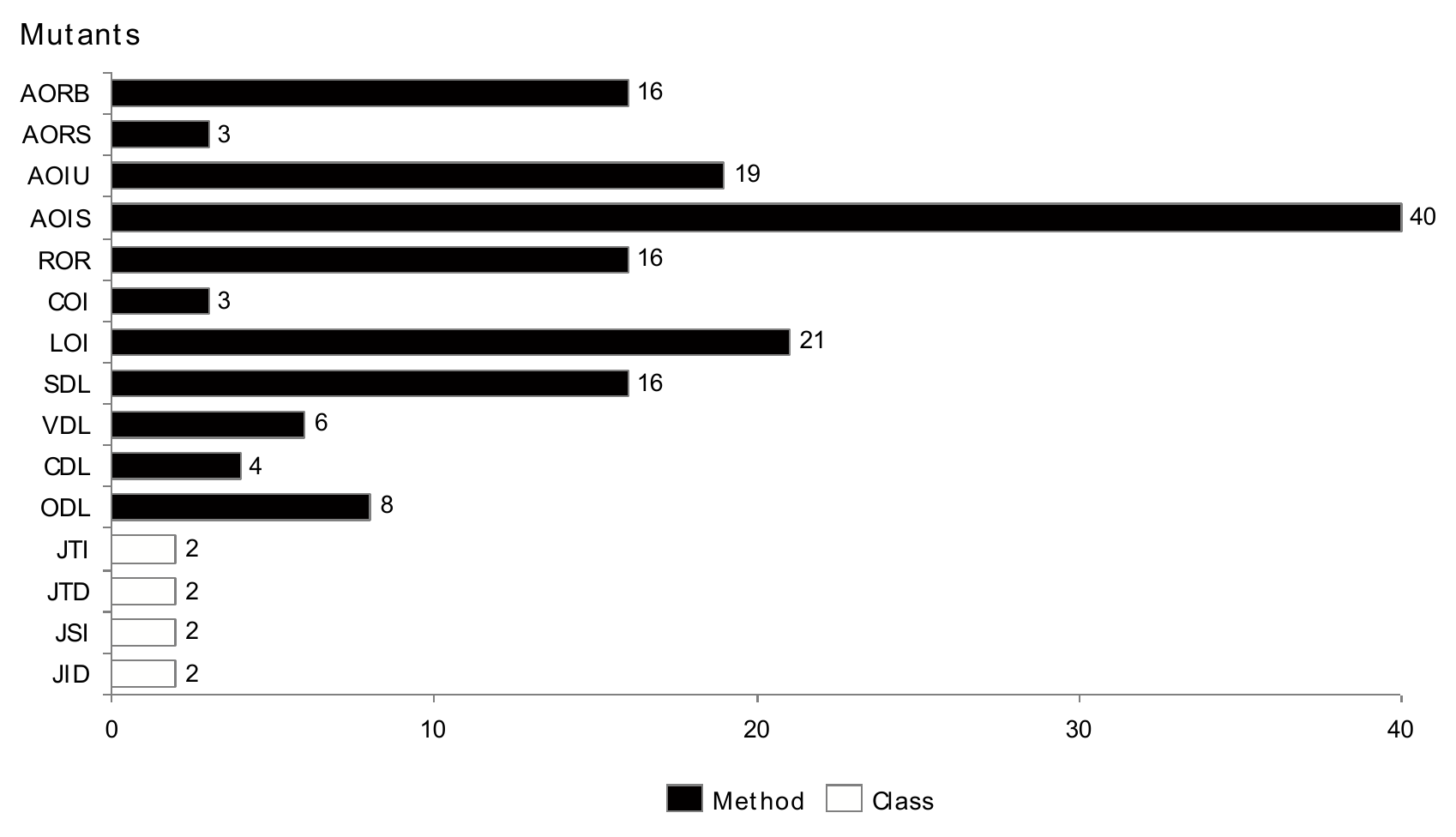}
\caption{Type and number of seeded faults into the Groovy program}
\label{GroovyFaultsSeeded}

\end{figure}

\begin{figure}
\centering
\includegraphics[width= 3.5 in]	{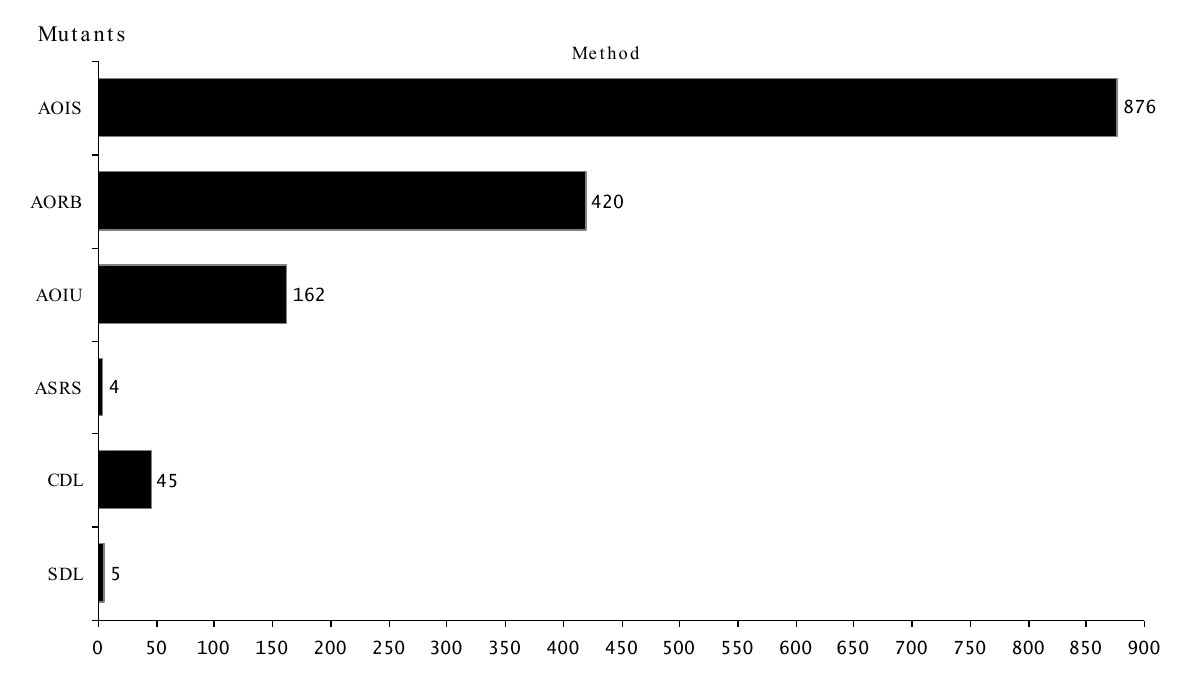}
\caption{Type and number of seeded faults into the Body calculator program}
\label{BoddyCalcultorFaultsSeeded}

\end{figure}

\begin{figure}
\centering
\includegraphics[width= 3.5 in]	{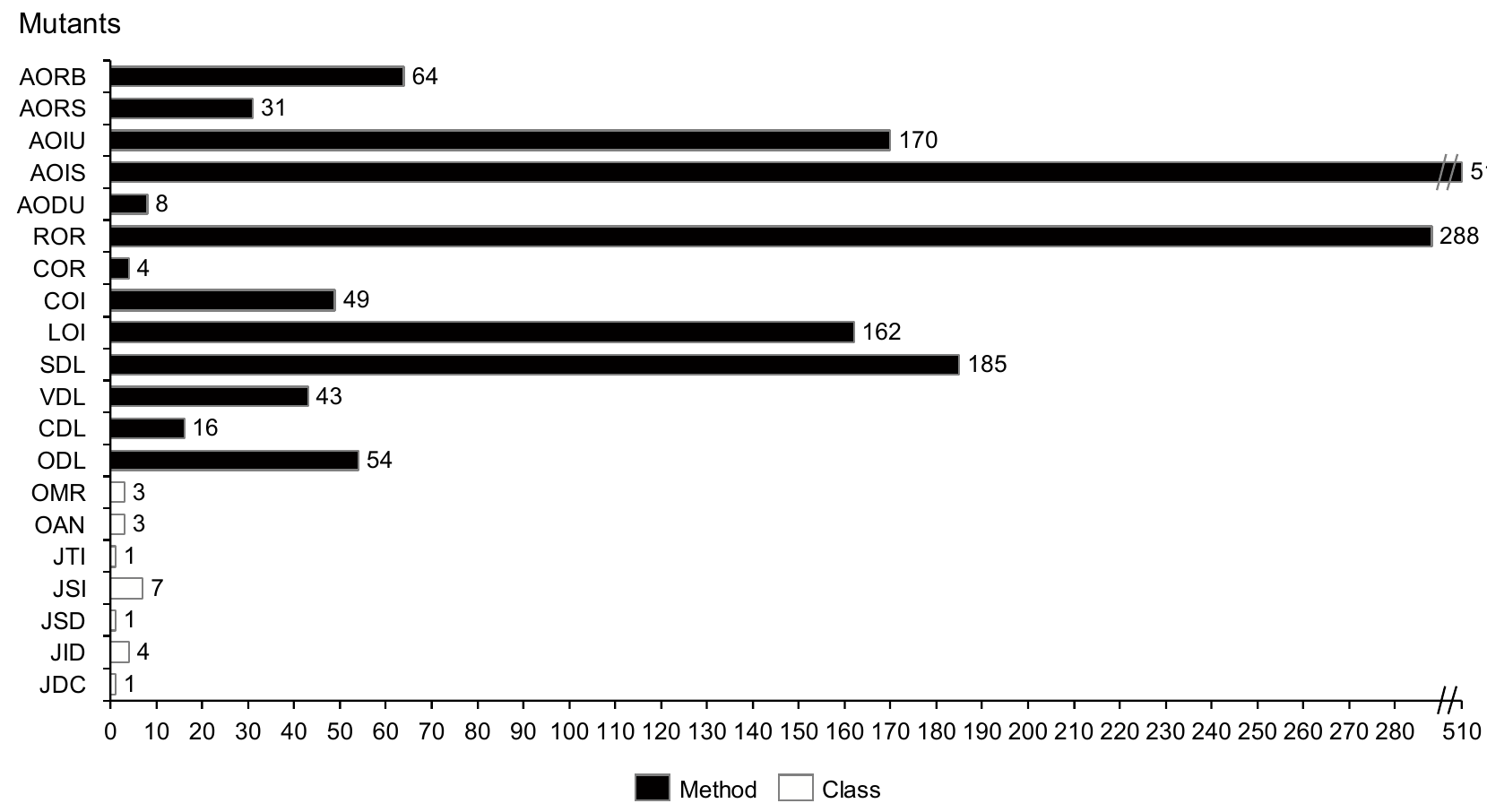}
\caption{Type and number of seeded faults into the Searching program}
\label{SearchFaultsSeeded}

\end{figure}

\begin{figure}
\centering
\includegraphics[width= 3.5 in]	{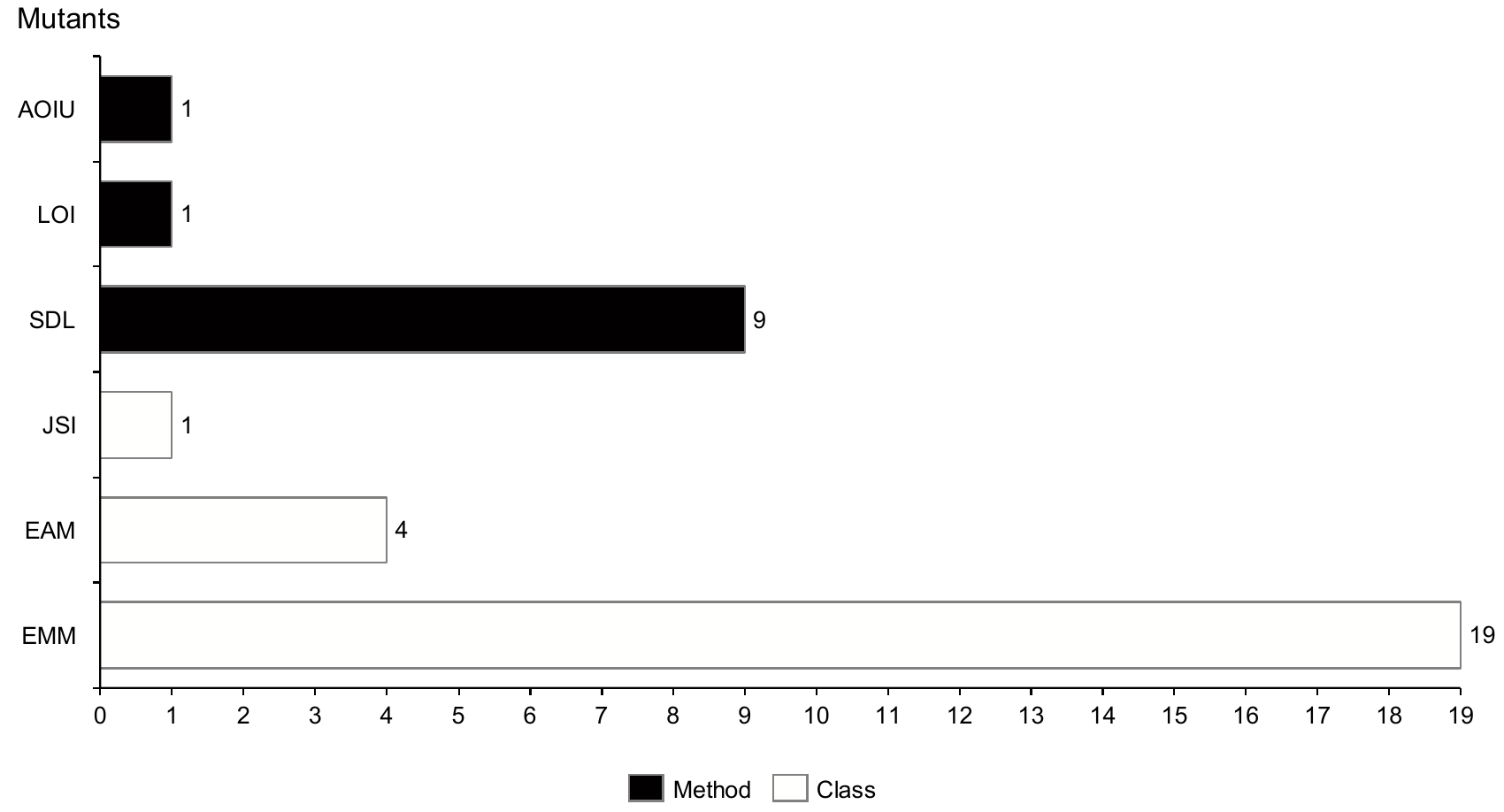}
\caption{Type and number of seeded faults into the Mortgage calculator program}
\label{MortgageFaultsSeeded}

\end{figure}

As previously mentioned, we have seeded many faults in each subjected program for the case study. Using $\mu java$, we have seeded 295 faults into the Replicated workers program, 160 faults into the Groovy program, 1512 faults into Body calculator program, 1584 faults into Searching program, and 35 faults into Mortgage program. Figures \ref{ReplicatedWorkerFaultsSeeded} - \ref{MortgageFaultsSeeded} shows the number and the type of all these faults for each subjected program individually.

%%%%%%%%%%%%%%% Bestoun 11/2/2019 --- 17

\subsection{Observations}

\subsubsection{RQ1. Parameter analysis - the effect on the internal code}\label{effectForValidity}

Based on the method in Section \ref{method}, we have analyzed each program. As illustrated in Figure \ref{Impact_Analysis}, we measured the code coverage of each input parameter in the testing framework. We changed the variable of the input parameter and measured the coverage to know the impact of the parameter on the internal code using Eq.\ref{Eq1}. For the case studies used in this paper, we double check the results also by manual inspection and reviewing the code of each program. Tables \ref{ReplicatedWorkParameterImpact} - \ref{MortgageParameterImpact} show the empirical results of the parameter impact analysis using the code coverage.

\begin{table*}

\caption{Replicated workers parameter impact analysis}
\label{ReplicatedWorkParameterImpact}
\centering{}%
\scriptsize
\begin{tabular}{|c|c|c|c|c|}
\hline 
Parameter Name & Type & Covered LOC & Uncovered LOC & Examples\tabularnewline
\hline 
\hline 
num\_workers & int & 30\% & 51\% & (1,2,5)\tabularnewline
\hline 
num\_items & int & 8\% & 73\% & (1,2,5)\tabularnewline
\hline 
min & float & 4\% & 77\% & (1,10,15)\tabularnewline
\hline 
max & float & 4\% & 77\% & (1,10,15)\tabularnewline
\hline 
epsilon & float & 81\% & 0\% & (0.05,0.1,0.25)\tabularnewline
\hline 
\end{tabular}
\end{table*}

\begin{table*}
\centering
\caption{Groovy parameter impact analysis}
\label{GroovyParameterImactAnalysis}
\scriptsize
\begin{tabular}{|l|c|c|c|c|}
\hline 
Parameter Name & Type & Covered LOC & Uncovered LOC & Examples\tabularnewline
\hline 
\hline 
threadCount & int & 35\% & 8\% & (0,1,2,8)\tabularnewline
\hline 
DEFAULT\_INITIAL\_CAPACITY & int & 5\% & 38\% & (0,1,100,1000)\tabularnewline
\hline 
DEFAULT\_LOAD\_FACTOR & float & 5\% & 38\% & (0,1,10,100)\tabularnewline
\hline 
MAXIMUM\_CAPACITY & int & 4\% & 39\% & (0,1,100,1000,10000)\tabularnewline
\hline 
concurrentReads & long & 6\% & 37\% & (0,1,10,1000)\tabularnewline
\hline 
\end{tabular}

\end{table*}

\begin{table*}

\caption{Body calculator parameter impact analysis}
\label{BodyParameterImpact}
\centering{}%
\scriptsize
\begin{tabular}{|c|c|c|c|c|}
\hline 
Parameter Name & Type & Covered LOC & Uncovered LOC & Examples\tabularnewline
\hline 
\hline 
male & boolean & 1\% & 32\% & true,false\tabularnewline
\hline 
age  & int & 2\% & 32\% & 0,10,20,50\tabularnewline
\hline 
weight  & int & 2\% & 31\% & 0,35,60\tabularnewline
\hline 
waist  & int & 1\% & 32\% & 0,30,50\tabularnewline
\hline 
hips  & int & 2\% & 31\% & 0,15,30\tabularnewline
\hline 
neck  & int & 16\% & 17\% & 0,100,200,400\tabularnewline
\hline 
heightArrayNum & int & 1\% & 32\% & 0,1,3,5\tabularnewline
\hline 
\end{tabular}

\end{table*}

\begin{table*}

\caption{Searching program parameter impact analysis}
\label{SearchingParameterImpact}
\begin{centering}
\scriptsize
\begin{tabular}{|c|c|c|c|c|}
\hline 
Parameter Name & Type & Covered LOC & Uncovered LOC & Examples\tabularnewline
\hline 
\hline 
Epsilon & double & 38\% & 15\% & 0.00,1.00,1E-8\tabularnewline
\hline 
minInt & int & 29\% & 24\% & 0,1,2,10\tabularnewline
\hline 
maxFrac & int & 28\% & 25\% & 0,1,2,10\tabularnewline
\hline 
minFrac & int & 28\% & 25\% & 0,1,2,10\tabularnewline
\hline 
size & int & 28\% & 25\% & 0,10,100\tabularnewline
\hline 
\end{tabular}
\par\end{centering}
\end{table*}

\begin{table*}

\caption{Mortgage parameter impact analysis}
\label{MortgageParameterImpact}
\begin{centering}
\scriptsize
\begin{tabular}{|c|c|c|c|c|}
\hline 
Parameter Name & Type & Covered LOC & Uncovered LOC & Examples\tabularnewline
\hline 
\hline 
mortgageAmout & int & 87\% & 3\% & 0,1,1000,1000000\tabularnewline
\hline 
mortgageTerm & int & 85\% & 5\% & 0,12,10000\tabularnewline
\hline 
interest & double & 85\% & 5\% & 0,0.10,1.00\tabularnewline
\hline 
startDate & int & 78\% & 12\% & 0,43500,90000\tabularnewline
\hline 
payment & double & 18\% & 72\% & 0,1,1000\tabularnewline
\hline 
extraMonthly & double & 7\% & 83\% & 0,1,1000\tabularnewline
\hline 
extraYearly & double & 5\% & 85\% & 0,1,10000\tabularnewline
\hline 
\end{tabular}
\par\end{centering}
\end{table*}

%%%%%%%%%%%%%%% Bestoun 11/2/2019 --- 18

Knowing the effect of each parameter on the program by trying all of its values gives a precise analysis of the sensitivity of the SUT by each one of them. Here, it is clear that not all the parameters have the same impact on the program. For example, the parameters "epsilon" and "num\_workers" have more impact on the replicated workers program as compared to other input parameters as they are covering more LOC. Similarly, the parameters "threadCount" and "neck" have more impact on the Groovy and body calculator programs respectively.

It should be mentioned here that the amount of coverage by each parameter gives the number of LOCs that is related only to that parameter and excluding the related LOCs to the other parameters. However, there are still common LOCs related to two or more parameter. Hence, we don't expect to have full coverage of LOCs by only one or even two input parameters. The full coverage of LOCs is not achievable even with all input parameters. This situation is due to different reasons such as poor programming and development practice of the SUT itself. 

%%%%%%%%%%%%%%% Bestoun 11/2/2019 --- 19

\subsubsection{RQ2. Correlation analysis - the input parameters' correlation}

Using the empirical results from parameter impact analysis, we can calculate the correlation among each tuple of input parameters. We used the standard correlation calculation in statistics in Eq. \ref{Eq:3} to know the interaction weight between the tuples as described in Section \ref{method}. Tables \ref{ReplicatedWorkersCorrelation} - \ref{MortgageCorrelation} show the correlation among the tuples of input parameters for each case study.

\begin{table*}

\caption{Replicated workers parameter correlation overview}
\label{ReplicatedWorkersCorrelation}
\centering{}%
\scriptsize
\begin{tabular}{|c|c|c|c|c|c|}
\hline 
Correlation & num\_workers & num\_items & min & max & epsilon\tabularnewline
\hline 
\hline 
num\_workers & - & 0.947 & 0.941 & 0.941 & 0.739\tabularnewline
\hline 
num\_items & 0.947 & - & 1.000 & 1.000 & 0.090\tabularnewline
\hline 
min & 0.941 & 1.000 & - & 1.000 & 0.047\tabularnewline
\hline 
max & 0.941 & 1.000 & 1.000 & - & 0.047\tabularnewline
\hline 
epsilon & 0.739 & 0.090 & 0.047 & 0.047 & -\tabularnewline
\hline 
\end{tabular}
\end{table*}

%%%%%%%%%%%%%%%%%%%%%%%%%%%%%%%%%%%%%%%%%%%

\begin{table*}
\caption{Groovy parameter correlation overview}
\centering
\label{GroovyCorrelation}

\includegraphics[width= 5 in]	{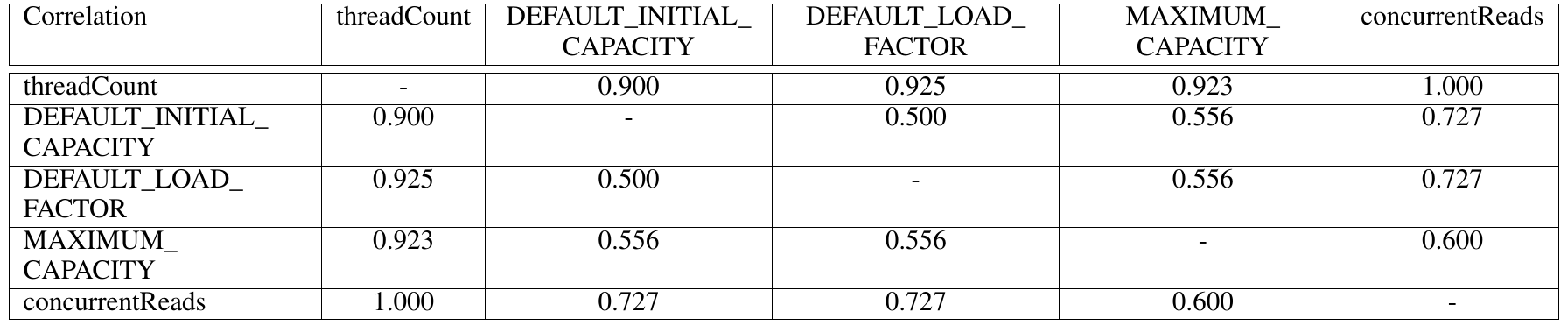}

\end{table*}

%%%%%%%%%%%%%%%%%%%%%%%%%%%%%%%%%%%%%

\begin{table*}

\caption{Body calculator parameter correlation overview}
\label{BodyCorrelation}
\centering{}%
\scriptsize
\begin{tabular}{|c|c|c|c|c|c|c|c|}
\hline 
Parameter & male & age  & weight  & waist & hips & neck & heightArrayNum\tabularnewline
\hline 
\hline 
male & - & 0.40 & 0.37  & 0.45  & 0.33  & 0.53  & 0.50 \tabularnewline
\hline 
age  & 0.40  & - & 0.94 & 0.37  & 0.29  & 0.51  & 0.80 \tabularnewline
\hline 
weight  & 0.37  & 0.94  & - & 0.34 & 0.27  & 0.51  & 0.74 \tabularnewline
\hline 
waist  & 0.45  & 0.37  & 0.34  & - & 0.97 & 0.58  & 0.45 \tabularnewline
\hline 
hips  & 0.33  & 0.29  & 0.27  & 0.97  & - & 0.50 & 0.33 \tabularnewline
\hline 
neck  & 0.53  & 0.51  & 0.51  & 0.58  & 0.50  & - & 0.53\tabularnewline
\hline 
heightArrayNum & 0.50 & 0.80 & 0.74 & 0.45 & 0.33 & 0.53 & -\tabularnewline
\hline 
\end{tabular}

\end{table*}

\begin{table}
\begin{centering}
\caption{Searching program parameter correlation overview}
\label{SearchingCorrelation}
\scriptsize
\begin{tabular}{|c|c|c|c|c|c|}
\hline 
Correlation & Epsilon & minInt & maxFrac & minFrac & size\tabularnewline
\hline 
\hline 
Epsilon & - & 0.58 & 0.61 & 0.61 & 0.68\tabularnewline
\hline 
minInt & 0.58 & - & 0.68 & 0.67 & 0.70\tabularnewline
\hline 
maxFrac & 0.61 & 0.68 & - & 0.50 & 0.71\tabularnewline
\hline 
minFrac & 0.61 & 0.67 & 0.50 & - & 0.71\tabularnewline
\hline 
size & 0.68 & 0.70 & 0.71 & 0.71 & -\tabularnewline
\hline 
\end{tabular}
\par\end{centering}
\end{table}

\begin{table*}

\caption{Mortgage program parameter correlation overview}
\label{MortgageCorrelation}
\scriptsize
\begin{centering}

\begin{tabular}{|c|c|c|c|c|c|c|c|}
\hline 
Correlation & mortgageAmout & mortgageTerm & interest & startDate & payment & extraMonthly & extraYearly\tabularnewline
\hline 
\hline 
mortgageAmout & - & 0.52 & 0.51 & 0.54 & 0.83 & 0.87 & 0.77\tabularnewline
\hline 
mortgageTerm & 0.52 & - & 0.51 & 0.52 & 0.81 & 0.87 & 0.80\tabularnewline
\hline 
interest & 0.51 & 0.51 & - & 0.52 & 0.78 & 0.85 & 0.79\tabularnewline
\hline 
startDate & 0.54 & 0.52 & 0.52 & - & 0.68 & 0.21 & 0.81\tabularnewline
\hline 
payment & 0.83 & 0.81 & 0.78 & 0.68 & - & 0.28 & 0.78\tabularnewline
\hline 
extraMonthly & 0.87 & 0.87 & 0.85 & 0.21 & 0.28 & - & 0.92\tabularnewline
\hline 
extraYearly & 0.77 & 0.80 & 0.79 & 0.81 & 0.78 & 0.92 & -\tabularnewline
\hline 
\end{tabular}
\par\end{centering}
\end{table*}

%%%%%%%%%%%%%%% Bestoun 11/2/2019 --- 20

Analyzing the correlation between the tuples of parameters gives a clear understanding of how each value of these parameters are related to each other and the strength of the relationship. We consider a relationship stronger as the value of the correlation approached to "1". Using this criterion for classification, we can categorize the tuples. For example, the parameters "min" and "num\_items" in the replicated workers program are highly related to each other as they have correlation value equal to 1. Similarly, the parameters "threadCount" and "concurrentReads" in the Groovy program are highly related to each other. In contrast, the interaction weight (i.e., relationship) between the "num\_items" and "epsilon" parameters is weak. As previously mentioned, we can put higher strength or full strength on those highly-related parameters and free the weakly-related parameters, i.e., marking them as "don't care" value in the test generation tool. Following this approach, we assure that those highly-related tuples will be fully covered in the test suite while the other tuples will appear in the test suite but not fully covered. 

%%%%%%%%%%%%%%% Bestoun 11/2/2019 --- 21

Table \ref{TestNumbers} shows the size of the generated test suite for each case study after and before the consideration of the impact analysis. It is noticeable that the size of the test suite decreased by a few test cases with the Replicated worker and the Groovy case studies. As we can see, for these two case studies there are a few parameters and not all of them are highly related to each other. As mentioned, during the test generation, we assigned "don't care" values to those parameters. As a result, the fewer (but effective) test cases were generated. Although it is not the aim of our study, this shows that in some cases our approach could also help to generate efficient test cases by generating fewer test cases.

As for the Body Calculator, searching, and Mortgage case studies, there is a higher number of parameters than the other two case studies. As we can see from the results, those parameters are related to each other. As a result, we considered different interaction strength during generation. Hence, the number of test cases goes higher.

\begin{table}

\caption{The size of the generated test suites before and after parameter impact consideration}
\label{TestNumbers}
\scriptsize
\centering{}%
\begin{tabular}{|l|c|c|}
\hline 
SUT & 2-way & mixed\tabularnewline
\hline 
\hline 
Replicated workers & 28 & 25\tabularnewline
\hline 
Groovy & 24 & 17\tabularnewline
\hline 
Body calculator & 19 & 62\tabularnewline
\hline 
Searching Program & 16 & 48\tabularnewline
\hline 
Mortgage program & 15 & 39\tabularnewline
\hline 
\end{tabular}

\end{table}

%%%%%%%%%%%%%%% Bestoun 11/2/2019 --- 22

\subsubsection{RQ3. Assessment of fault detection - effectiveness of the new approach}

\begin{figure}
\centering
\includegraphics[width= 4 in]	{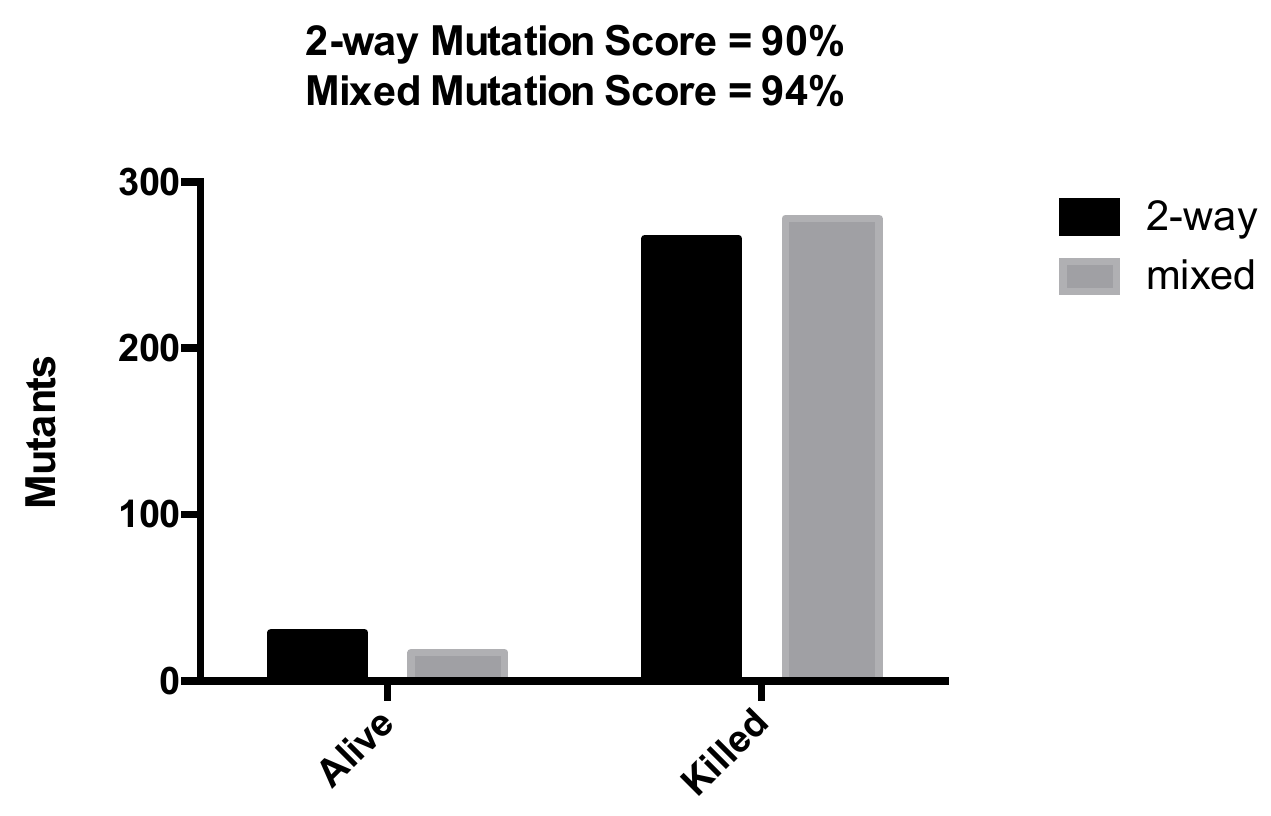}
\caption{Comparison of the test suite effectiveness for detecting faults in case of Replicated worker program}
\label{ReplicatedWorkResults}

\end{figure}

\begin{figure}
\centering
\includegraphics[width= 3 in]	{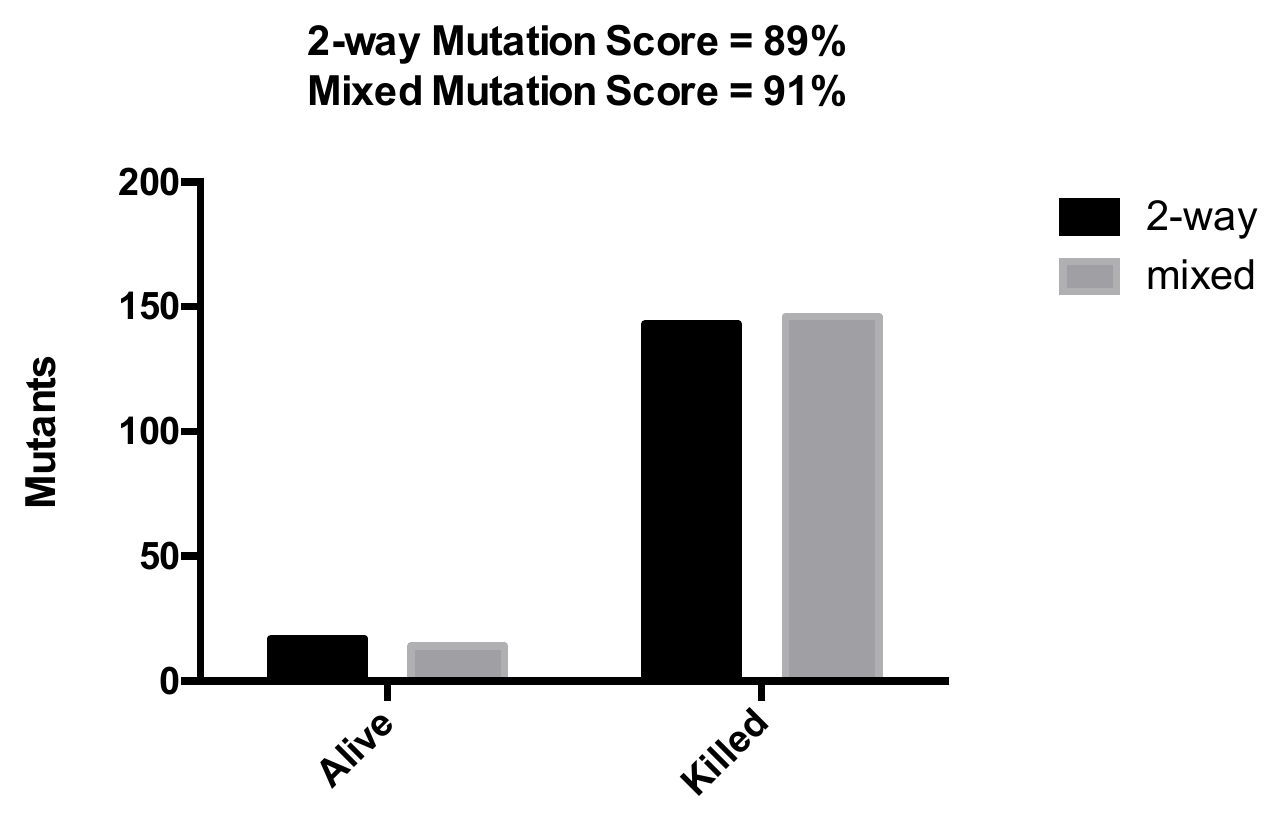}
\caption{Comparison of the test suite effectiveness for detecting faults in case of Groovy program}
\label{GroovyResults}

\end{figure}

\begin{figure}
\centering
\includegraphics[width= 3 in]	{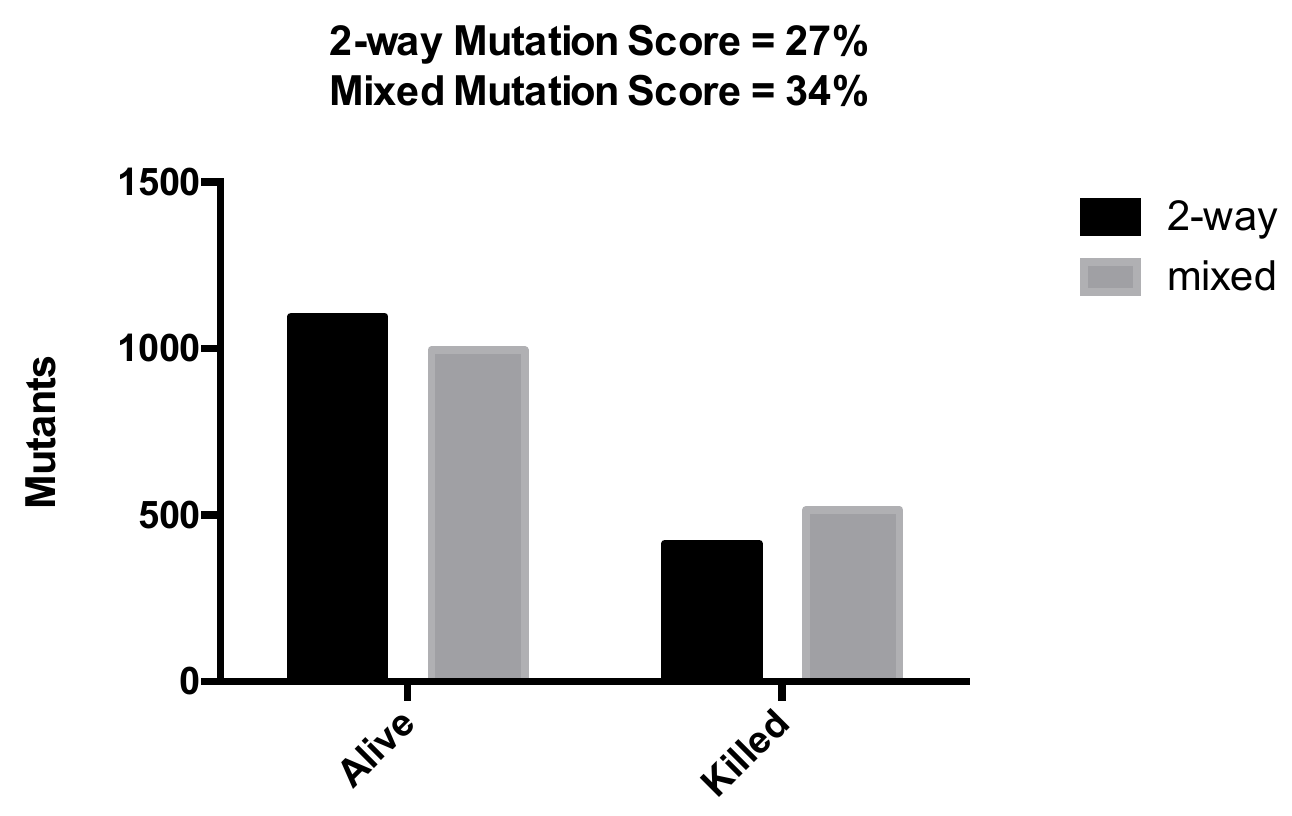}
\caption{Comparison of the test suite effectiveness for detecting faults in case of Body calculator program}
\label{BodycalculatorResults}

\end{figure}

\begin{figure}
\centering
\includegraphics[width= 3 in]	{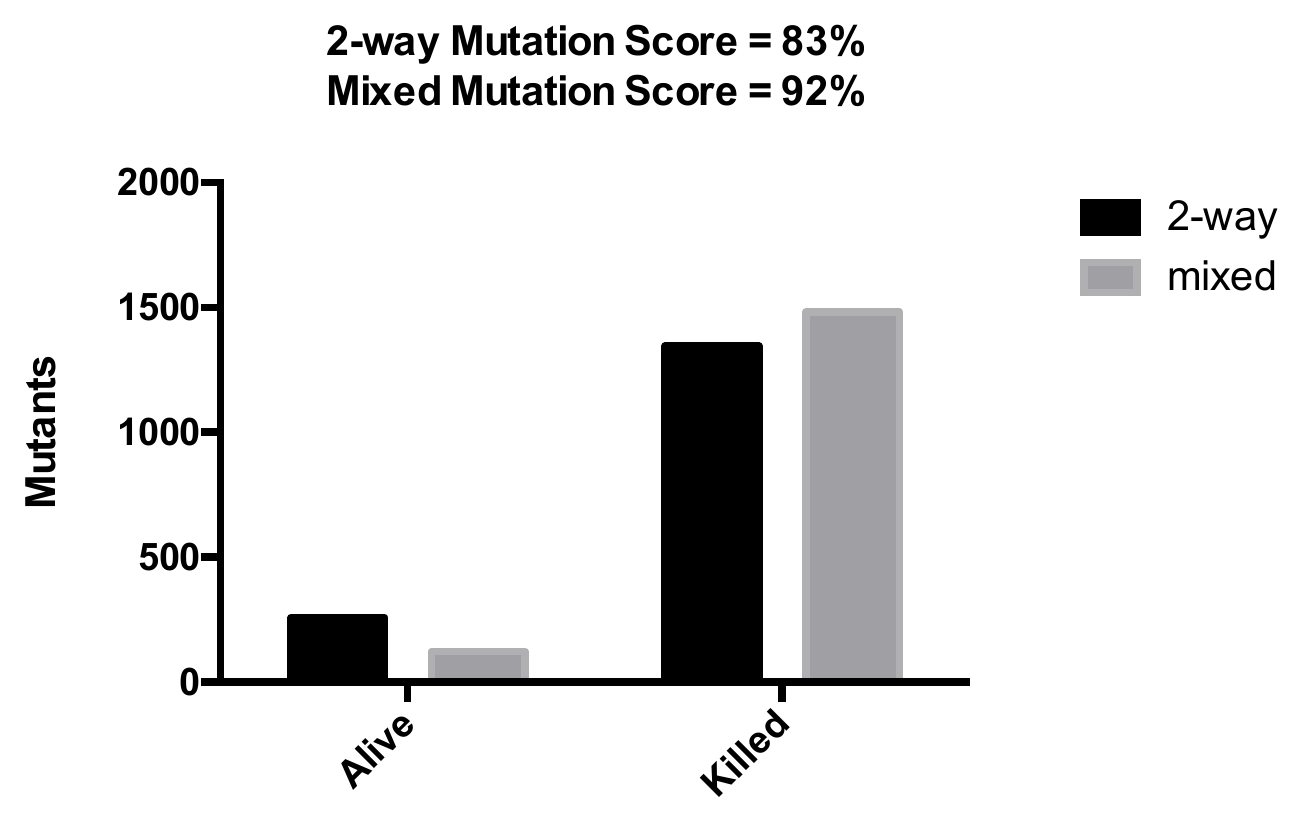}
\caption{Comparison of the test suite effectiveness for detecting faults in case of Searching program}
\label{SearchingMutationResults}

\end{figure}

\begin{figure}
\centering
\includegraphics[width= 3 in]	{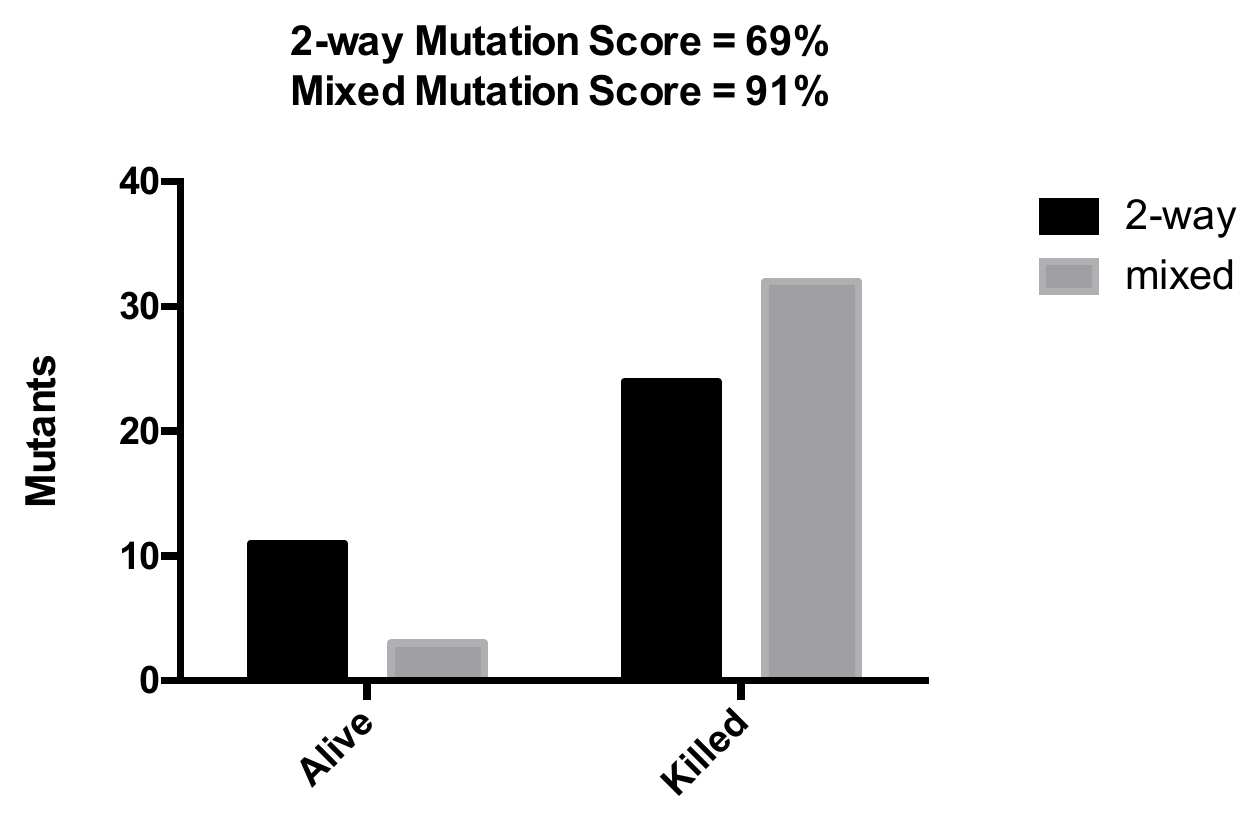}
\caption{Comparison of the test suite effectiveness for detecting faults in case of Mortgage program}
\label{MortgageMutationResults}

\end{figure}

As mentioned previously, we have injected different mutants into each program for possible fault detection. The experiments aim to know the effectiveness of our approach to detect new faults as compared to the classical black-box CIT. Figures \ref{ReplicatedWorkResults} - \ref{MortgageMutationResults} show that, in general, the CIT is an effective approach to detect faults. The figures also show that using our approach (mixed strength), we can detect more faults as compared to pairwise (i.e., 2-way) testing. Here, the code-aware CIT can detect more faults by giving interaction strength to those highly correlated parameters that have been measured during the analysis process at the first stage.

As in the case of the replicated worker program in Figure \ref{ReplicatedWorkResults}, it is clear that there is a 4\% higher mutation score as compared to the 2-way test suite. Similarly, we can see the results of the Groovy program in Figure \ref{GroovyResults}. Here, with the new approach, we can detect 16 more faults. Moreover, we can see in Figure \ref{BodycalculatorResults} that more faults can be detected in the case of the Body calculator program. Here, 138 more faults were detected as compared to the 2-way test suite. As for the case of Searching and Mortgage programs in Figures \ref{SearchingMutationResults} and \ref{MortgageMutationResults}, more faults detected in with the mixed strength interaction that leads to a better mutation score. Specifically, for the Searching program, 136 new faults were discovered that leads to a 9\% better mutation score as compared to the 2-way test suite. Similarly, for the Mortgage program, eight new faults were discovered by the mixed strength test suite, that leads to a 22\% better mutation score.

%%%%%%%%%%%%%%% Bestoun 11/2/2019 --- 23

\begin{figure}
\centering
\includegraphics[width= 3 in]	{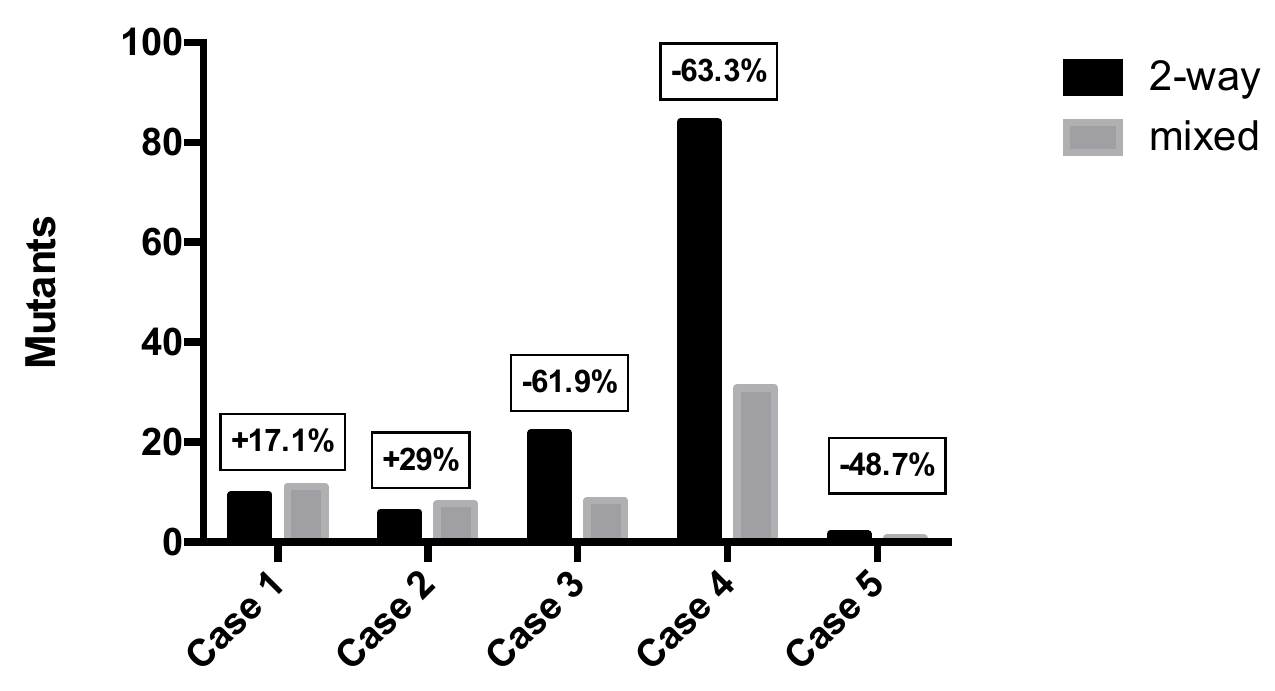}
\caption{Overall Efficiency of all the five case studies}
\label{OverallPeformance}

\end{figure}

We note that there are still many faults alive and our test suites have not killed them. Practically, it is not possible to kill all the mutants with a combinatorial interaction test suite. Those faults may be killed using some other test generation algorithms as they may not be interaction faults. The alive faults may also be detected by combinatorial interaction test suites when the interaction strength is greater than two (i.e., $t>2$). In this paper, we only aimed to show the effectiveness of the new approach. As far as the approach is effective, it can be used with higher strength easily. 
%%%%%%%%%%%%%%% Bestoun 11/2/2019 --- 24

If we combine the results from all five case studies, we get an impressive discussion. As can be seen in  Figure \ref{OverallPeformance}, the code-aware CIT performed better. On average it is better by 3\% in mutant detection and by 25\% in efficiency (Calculated by detected mutations divided by the number of test cases). This is however largely variable based on the nature of the program. In all five cases, the code-aware CIT managed to find new mutants compared to the reference pairwise test suite. The specific number is tied to the total lines of code/mutants used, but in the first and second case studies, there is an efficiency gain of around 17\% and 29\%, while there is no gain in the last three case studies. However, as shown previously, there are still several new killed mutants that makes our approach effective. In fact, the efficiency is computed by dividing the number of mutants killed by the number of test cases in the test suite. For example, in the case of Body calculator, the input data is 415/1512 = 27\% rounded for pairwise and for 516/1512 = 34\% for the mixed test suite. The corresponding efficiency is thus 415/19=21.84 and 516/62=8.32. Here, the -62\% comes from the relative change between the two efficiencies, i.e., $8.32-21.84)/21.8$.

\section{Threats to Validity}\label{threatsToValidity}

As in other empirical studies, our study is subjected to validity threats. We have tried to eliminate these threats during our experiments. We have redesigned the experiments several times to avoid different threats to validity. However, for the sake of reliability, we outline some a few significant threats that we have faced during our experiments. 

Regarding the generalization of our results (i.e., external validity), we have studies only five case studies written in Java, and different results may be reported for other programs. We used more than one program to validate our approach. Also, we tried to avoid hand-generated seeded faults by using a standard Java mutation tool to assure reliability. The faults injected by the $\mu java$ are more realistic faults than the hand-generated faults.

Regarding the effect of other internal factors on our results (i.e., internal validity), there might be other factors responsible for these results that we obtain due to the instrumentation of the case studies' code. However, we have tried to run the program with and without the instrumentation and also we have double checking our experiments and manually reviewing the codes and the obtained results for a few cases. Also, as we illustrated in Section \ref{effectForValidity}, internally, the parameters may affect each other, and there could be a threat that the effects of parameters were overlapping. We have tried to eliminate this threat by changing the values of the parameters systematically one by. Hence, we assure a fair measure even if it is not presenting the actual effect of that parameter on the code. 

%%%%%%%%%%%%%%% Bestoun 11/2/2019 --- 25

\section{Conclusion}\label{Conclusion}

In this paper, we have presented our new code-aware approach for conducting the combinatorial interaction testing. We showed the results of an empirical study to examine the effectiveness of our approach through five case studies. We first examined the SUT by using a code coverage framework to analyze the impact of each input parameter. Then, we used the correlation coefficient to assess the relationship of the input parameters to each other. Using these assessment and analysis steps, we were able to reconsider the generated test cases by taking those correlated parameters into account. We aim to study this approach experimentally. Although we have automated many steps during our experiments, developing an automated tool to conduct this approach is beyond the scope of this paper. Nevertheless, we have demonstrated an algorithm pseudo-code that can be automated through a tool implementation in the future. This paper could serve as a strong base for a future research direction to develop an automated testing generation tool for the code-aware CIT.  

As we can see from the experiments, using the parameter impact analysis can be utilized to generate effective test suites for fault detection. We named this testing process code-aware CIT. Although the fault detection rates and mutation score may vary from a program to another, the results showed that this approach is worth pursuing as another variant of CIT for practical aspects. The results also showed that this approach could be effective to find new faults that cannot be detected by the traditional t-wise testing. Hence, this approach can be treated as a complementary not a substitution of the t-wise testing.

In addition to the directions mentioned above, there are many other future directions of code-aware CIT. Examining the effect of the input parameters on the internal code and using this effect as a relationship to generate more effective test cases is essential. One possible direction is to study the impact of the input sequence and sequence-less on the code and the generation process. Using the data flow in the code level and the interaction direction could also be an essential study finding. Shi et al. \cite{Shi2012} demonstrated that there is some interest in analyzing the interaction direction at the code level. This could also affect the generation method of the test suite.

\bibliographystyle{model1-num-names}
\bibliography{sample}
\end{document}